\documentclass[prl,superscriptaddress,amsfonts,amssymb,amsmath,floats,twocolumn,aps,footinbib,shownopacs]{revtex4-2}
\usepackage[pdftex]{graphicx}
\usepackage{subfigure}
\usepackage{dsfont}
\usepackage{lipsum}
\usepackage{dcolumn}
\usepackage{bm}
\usepackage{color}
\usepackage{physics}
\usepackage{multirow}
\usepackage{cancel}
\usepackage[pdftex,colorlinks=true, linkcolor=blue,citecolor=blue,filecolor=blue]{hyperref}
\newcommand{\bb}[1]{\mathbf{#1}}
\newcommand{\m}[1]{\mathcal{#1}}

\newcommand{\am}[1]{\langle \m{#1} \rangle}

\begin{document}

\title{Nonequilibrium transport through an interacting monitored quantum dot} 

\author{Daniel Werner}
\affiliation{Institute of Theoretical and Computational Physics, Graz University of Technology, 8010 Graz, Austria}
\author{Matthieu Vanhoecke}
\affiliation{JEIP, UAR 3573 CNRS, Coll\`ege de France, PSL Research University, 11 Place Marcelin Berthelot, 75321 Paris Cedex 05, France}
\author{Marco Schir\`o}
\affiliation{JEIP, UAR 3573 CNRS, Coll\`ege de France, PSL Research University, 11 Place Marcelin Berthelot, 75321 Paris Cedex 05, France}
\author{Enrico Arrigoni}
\affiliation{Institute of Theoretical and Computational Physics, Graz University of Technology, 8010 Graz, Austria}

\begin{abstract}
We study the interplay between strong correlations and Markovian dephasing, resulting from monitoring the charge or spin degrees of freedom of a quantum dot described by a dissipative Anderson impurity model. Using the Auxiliary master equation approach we compute the steady-state spectral function and occupation of the dot and discuss the role of dephasing on Kondo physics. Furthermore, we consider a two-lead setup which allows to compute the steady-state current and conductance. We show that the Kondo steady-state is robust to moderate charge dephasing but not to spin-dephasing, which we interpret in terms of dephasing-induced heating of low-energy excitations.  Finally, we show universal scaling collapse of the non-linear conductance with a dephasing-dependent Kondo scale. 
\end{abstract}
\maketitle 	

\emph{Introduction - } Traditionally considered as detrimental for quantum coherence, dissipation due to coupling to an external environment has recently emerged as a powerful tool to control and prepare quantum systems~\cite{verstraete2008quantum,harrington2022engineered}. For example, tailored dissipative processes are currently used to extend the lifetime of superconducting qubits~\cite{lescanne2020exponential,reglade2024quantum}. Similarly, continuous monitoring and quantum measurements offer a non-unitary way to prepare and manipulate quantum systems~\cite{buffoni2019quantum,barontini2025quantum}. These breakthroughs in the field of quantum information and computation have raised interest on the possibility of manipulating many-body quantum systems via dissipative processes~\cite{ma2019dissipatively,mi2023stable,google_dephasing_abanin,langbehn2024dilute,fazio2025manybodyopenquantumsystems}. 

Mesoscopic systems provide a natural setting where to explore these ideas. Quantum transport through correlated quantum dots has been the source of exciting many-body physics over the past decades~\cite{Kouwenhoven_2001,Pustilnik_2004}. The Kondo effect~\cite{kondo1964resistance,anderson1961localized,wilson1975therenormalization} and its different incarnation and generalization in and out of equilibrium for example provide a testbed for theories of strongly correlated electron systems~\cite{hewson1993thekondo}. Coupling quantum dots with controlled dissipative environments such as charge detectors~\cite{wiseman2009quantummeasurementand,Tilloy_2014,landi2024current} or microwave cavities~\cite{delbecq2011coupling} has been demonstrated, with their impact on Kondo physics starting to be investigated~\cite{desjardin2017observation}. On a different experimental front, transport via quantum impurities with ultra-cold atoms has also made tremendous progress in recent years~\cite{barontini2013controlling,riegger2018localized,lebrat2019quantized,nagy2018exploring,huang2023superfluid}. These settings naturally include fast dissipative processes such as dephasing due to stimulated emission~\cite{gerbier2010heating,bouganne2020} or correlated particle losses~\cite{garcia-ripoll2009,meausurement2015patil,TomitaEtAlScienceAdv17,honda2023observation}.

Theoretical investigations on dissipative quantum impurity models, where localised levels coupled to a bath are exposed to fast Markovian dissipation, have recently seen a blooming~\cite{Scarlatella2019,Froml2019,Tonielli2019,visuri2022symmetry,ferreira2023exact,stefanini2023orthogonality,stefanini2024dissipative,vanhoecke2024diagrammatic,vanhoecke2025dissipativekondophysicsanderson,qu2025variational}.  In particular, the interplay between correlations and dissipation due to dephasing or equivalently monitoring of the dot charge has been shown to give rise to Kondo-Zeno crossover in the dynamics ~\cite{vanhoecke2025kondozenocrossoverdynamicsmonitored}. A crucial question concerns the interplay between Kondo physics and dissipation in the steady-state transport, which is of direct experimental relevance. Transport through a non-interacting monitored dot has been recently shown to contain rich physics~\cite{ferreira2023exact} and it is therefore natural to investigate the interacting case. At more fundamental level one could ask what is the effect of dephasing and monitoring on Kondo temperature and whether the former acts as an infrared cut-off like temperature or bias voltage for the conventional (unitary) Kondo problem~\cite{rosch2001kondo}. 
\begin{figure}[!t] 
 \includegraphics[width=0.5\textwidth]{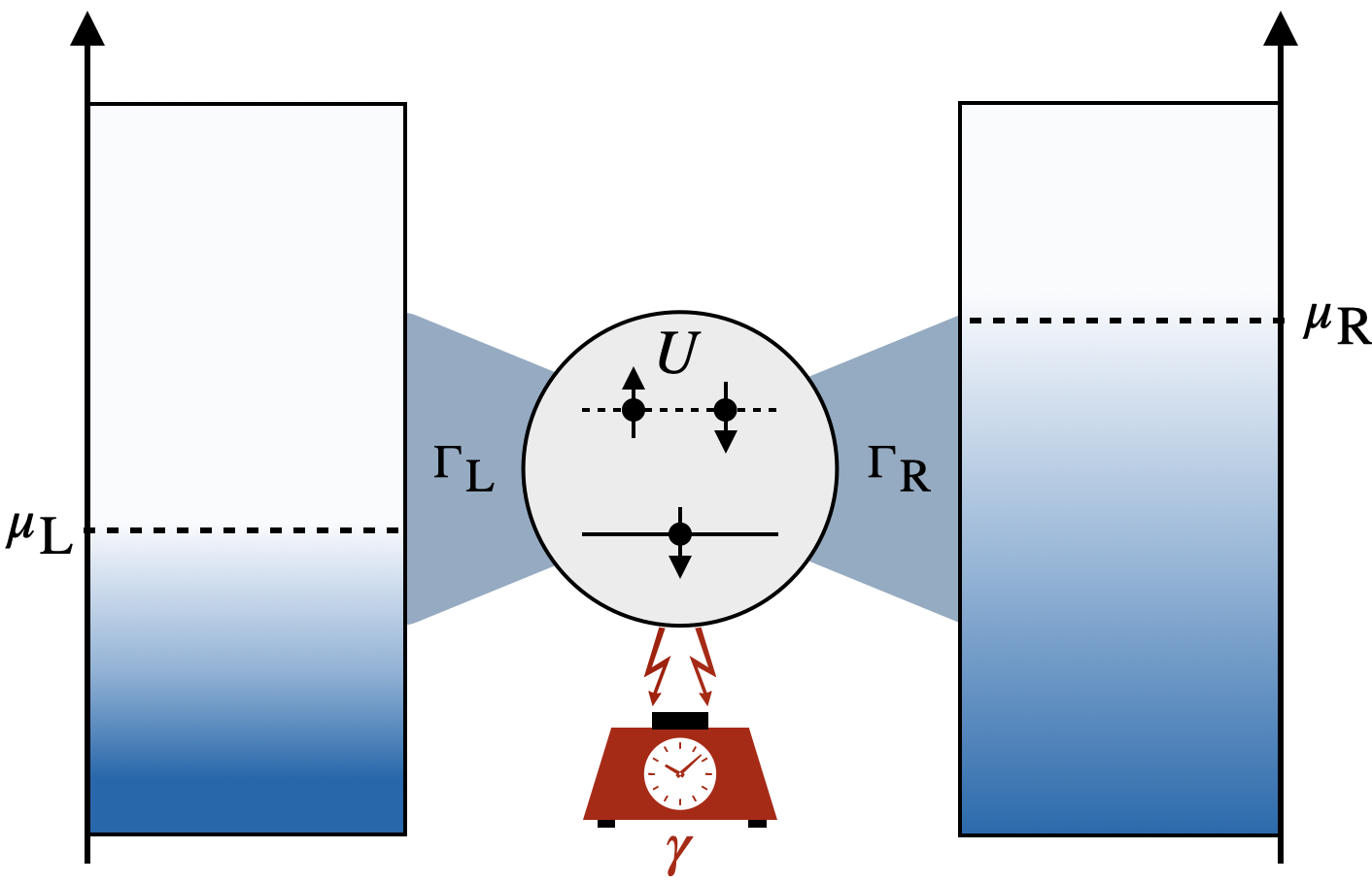}
    \caption{\label{fig:sketch}
 Sketch of the setup for the two leads dissipative Anderson Impurity Model: a quantum dot subjected to Markovian dissipation due to continuous monitoring with strength $\gamma$ and coupled to two large metallic leads via a hybridization $\Gamma_{\rm R,L}$. The two leads are held at different chemical potential leading to a current flowing and nonequilibrium transport.}
\end{figure}

In this Letter we study the steady-state transport via an interacting monitored quantum dot, described by a dissipative Anderson impurity model with charge or spin dephasing. Using the Auxiliary master equation approach~\cite{do.nu.14,ar.do.18,we.lo.23} we compute the steady-state dot Green's function and from these obtain linear and non-linear conductance as a function of the monitoring rate and the interaction. We show that weak charge dephasing leaves visible signatures of Kondo physics, both in the spectral function and in the conductance, as opposed to acting on the spin sector which rapidly destroys many-body coherent effects. We interpret this result in terms of effective heating of low-energy degrees of freedom which is enhanced when the spin rather than the charge of the dot is monitored. Finally we show that the non-linear conductance shows a perfect scaling collapse with a dephasing dependent scale, suggesting that Kondo universality survives in presence of dissipation.

\begin{figure*}[t!] 
    \centering
    \includegraphics[width=1.0\textwidth]{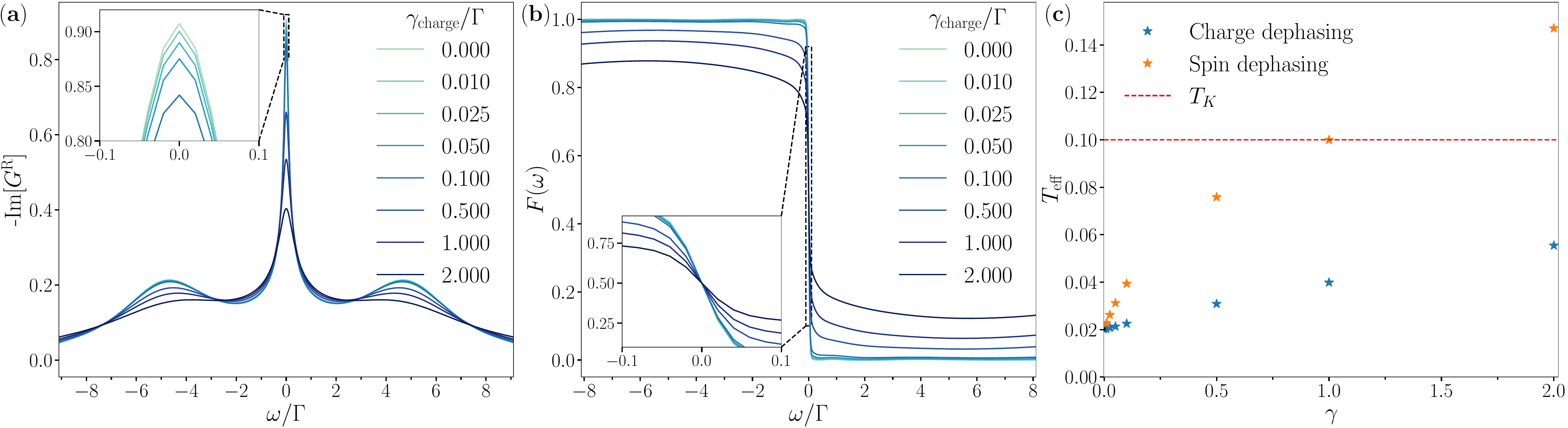}
    \caption{\label{fig:spectralfunction}  Dissipative Anderson Impurity Model - (a) Impurity spectral function at half-filling for interaction strength 
$U = 8\Gamma$, $V_g=0$, shown for increasing values of the charge dephasing rate $\gamma_{\rm charge}$.
    (b) Corresponding distribution function $F(\omega)$ as $\gamma_{\rm charge}$ increases.
    (c) Effective temperature $T_{\rm eff}$ extracted from the distribution function, as a function of $\gamma$ in the charge dephasing case, compared with the spin dephasing case and the Kondo temperature of the non-dissipative Anderson impurity model (taken from Ref.~\cite{we.lo.23}).
    The leads are held at a finite temperature $T=0.02 \Gamma$, and are modeled with a flat (wide-band) density of states. 
    }
\end{figure*}
\emph{Dissipative Anderson Impurity Model - } We consider a model for an interacting spinful quantum dot coupled to two metallic reservoirs at different chemical potential (see sketch in Fig.~\ref{fig:sketch}), as described by the Anderson Impurity Model (AIM)~\cite{anderson1961localized,hewson1993thekondo} with Hamiltonian
\begin{align}
    \mathcal{H} = \mathcal{H}_{\rm leads}  + \mathcal{H}_{\rm hyb} + \mathcal{H}_{\rm dot} \,.
\end{align}
Here the first term describes the Hamiltonian of the non-interacting leads $\mathcal{H}_{\rm leads}=\sum_{\sigma, \mathbf{k} \alpha} \epsilon_{\mathbf{k}\alpha}\, c_{\mathbf{k},\sigma \alpha}^\dagger c_{\mathbf{k},\sigma \alpha}$, where $c^\dagger_{\mathbf{k},\sigma \alpha},c_{\mathbf{k},\sigma \alpha}$ are the fermionic operators operators of the lead $\alpha=L,R$, with momentum $\mathbf{k}$, and spin $\sigma$. The lead are characterized by the single-particle energy
$\epsilon_{\mathbf{k}\alpha}=\epsilon_{\mathbf{k}}-\mu_{\alpha}$, where the chemical potentials of the leads are set by the applied bias voltage $\mu_{L}-\mu_R=eV$. the second term describes the hybridization between dot and
the leads $\mathcal{H}_{\rm hyb} = \sum_{\mathbf{k},\sigma \alpha } \left( V_{\mathbf{k} \alpha}\, d_\sigma^\dagger c_{\mathbf{k},\sigma \alpha} + V^{*}_{\mathbf{k} \alpha}\, c_{\mathbf{k},\sigma \alpha}^\dagger d_\sigma \right)$ with coupling $V_{\bb{k} \alpha}$. Finally, the last term describes the dot Hamiltonian $\mathcal{H}_{\rm dot}=\sum_\sigma \left(V_g-\frac{U}{2}\right)\, d_\sigma^\dagger d_\sigma + U\, n_\uparrow n_\downarrow$, where $d^\dagger_\sigma$ and $d_\sigma$ are the dot creation/annihilation operators, $V_g$ is the gate voltage 
and $U$ the Coulomb interaction, with $n_\sigma = d^\dagger_\sigma d_\sigma$. In the following, we consider the fermionic leads to be at finite temperature $T$ and assume them to have a flat (constant) density of states
 $\Gamma_{\alpha}(\epsilon) = \pi \sum_{\bb{k}} \vert V_{\bb{k} \alpha} \vert^2 \delta(\epsilon - \epsilon_{\bb{k}\alpha}) = \Gamma_{\alpha} \Theta_{sm}(D-\vert \epsilon \vert) $, where $D$ is the half-bandwidth of the lead~\footnote{
Here $\Theta_{sm}$ is a $\Theta$ function which is smoothed at the edges in order to avoid sharp features.} Although simplistic this density of states encodes the main properties of a metallic conduction lead, with a finite bandwidth and a finite weight at the Fermi level. 

In addition to coherent processes on the dot described by $\mathcal{H}_{\rm dot}$ we are interested in a situation where the impurity is exposed to local dissipative processes, that we assume to be Markovian and described by a Lindblad Master Equation~\cite{breuerPetruccione2010}. This dissipative processes originate from some fast Markovian environment, whose microscopic degrees of freedom are not under our control and so can be traced
out from the start. This has to be contrasted with the quantum bath described by the fermions $c_{\bb{k},\sigma \alpha},c_{\bb{k},\sigma \alpha}^\dagger$ which play a key role in the many-body physics of the quantum impurity. As a result of this local dissipation the entire system (metallic leads plus quantum dot) is described by a density matrix $\rho_t$ which evolves in time according to the Lindblad equation
\begin{align}
    \partial_t \rho_t = -i \left[ H, \rho_t \right] +  L_b \rho_t L_b^\dagger - \frac{1}{2}\{L_b^\dagger  L_b , \rho_t  \}
\end{align}
Where here we have denoted $L_b , L_b^\dagger$ the jump operators acting on  the
impurity system only, that is to say they are written only as functions of the operators $d_\sigma,d_\sigma^\dagger$, and $b = \{\rm{charge,spin} \} $ denotes the operator that is being monitored. In this work, we consider the case of continuous monitoring of some hermitian operator of the dot which leads to dissipation in the form of dephasing. 
In particular we will consider separately dephasing of the total charge, i.e. a jump operator of the form 
\begin{align}
L_{\rm charge} = \sqrt{\gamma_{\rm charge}} \sum_\sigma  n_\sigma
\end{align}
and of the spin along the $z-$ axis, 
\begin{align}
L_{\rm spin} = \sqrt{\gamma_{\rm spin}} \sum_\sigma \sigma n_\sigma
\end{align}
We note that these jump operators commute with the total particle number, i.e. they describe energy rather than particle exchange with the Markovian bath, and furthermore preserve particle-hole symmetry of the Lindblad maser equation, which is therefore guaranteed for the neutrality point $V_g=U/2$.

Our results are obtained using the Auxiliary Master Equation Approach (AMEA)\cite{ar.kn.13,do.nu.14,we.lo.23,supplementary}. In AMEA, the metallic leads are replaced by an auxiliary environment composed of a finite number $N_B$ of bath sites and additional Markovian reservoirs. The resulting many-body Lindblad equation, which fully accounts for electronic correlations on the impurity, can be solved exactly using numerical methods. AMEA is a nonperturbative approach, and its accuracy is governed solely by the deviation between the hybridization functions of the original leads and those of the auxiliary baths. This deviation decreases exponentially with $N_B$, as demonstrated in Ref.\cite{do.so.17}.
In this work, we solve the auxiliary Lindblad problem using the Configuration Interaction (CI) scheme introduced in Ref.\cite{we.lo.23}, which enables access to the Kondo scaling regime of the conductance for interaction strengths up to $U/\Gamma\sim 8$. A key advantage of AMEA is its equal applicability to both equilibrium and nonequilibrium impurity problems producing real frequency spectra. For the present case, AMEA is particularly well suited: the inclusion of the Lindblad dissipative term does not incur any additional computational cost. In the Supplementary Material\cite{supplementary}, we benchmark AMEA against the exact solution for a non-interacting dissipative impurity ($U=0$) and find perfect agreement.

\emph{Spectral and Distribution Function of the dot} -  We start our discussion from the spectral function of the dot, which is a key quantity to characterize Kondo effect and it is defined as
\begin{align}
    A(\omega) = -\frac{1}{\pi} \rm{Im}\left[ G^{R}(\omega)\right]
\end{align}
where in the time domain the retarded Green's function 
is defined as,
\begin{align}
    G^{R}(t,t^\prime) = -i \theta(t-t^\prime) \langle \{d_\sigma(t),d^\dagger_\sigma(t^\prime)\} \rangle
\end{align}
We plot the spectral function in Fig.~\ref{fig:spectralfunction}(a) for a fixed temperature $T=0.02\Gamma$ and $U=8\Gamma$, well within the Kondo regime ($T_K=0.1\Gamma$ for these parameters, see dashed line in panel (c)) and increasing values of $\gamma_{\rm charge}$. We see that charge dephasing reduces the height of the Kondo peak, which is expected since dissipative impurities lack the equivalent of Luttinger theorem~\cite{vanhoecke2025kondozenocrossoverdynamicsmonitored}, but does not broaden the resonance at least up to values of $\gamma_{\rm charge}\sim \Gamma$, well above $T_K$. The incoherent excitations at high-frequency, corresponding to the Hubbard bands are suppressed already for $\gamma_{\rm charge}<U/2$. This supports the indication that Kondo effect is robust to moderate charge dephasing. For comparison we show in Ref.~\cite{supplementary} that spin dephasing suppresses the Kondo resonance already for $\gamma_{\rm spin}\ll T_K$, as expected for a relevant perturbation of the Kondo fixed point. To understand better the difference between the two dissipative processes we look at the Keldysh Green's function defined as,
\begin{align}
    G^{K}_{\sigma}(t,t^\prime ) = -i \langle  \left[ d_{\sigma}(t), d_\sigma^\dagger (t^\prime )\right]\rangle 
\end{align}
By expressing the Keldysh Green's function in frequency space and relating it to the spectral function, one can write
\begin{align}
    G^{K}(\omega) = \left(1 -2 F(\omega) \right)  \left[ G^{R}_{\sigma}(\omega) - G^{A}_{\sigma}(\omega) \right],
\end{align}
where \( F(\omega) \) is the distribution function. In Fig.~\ref{fig:spectralfunction}(b), we plot the distribution function for increasing values of $\gamma_{\rm charge}$. We see that the main effect of charge dephasing is to suppress uniformly the occupation of finite frequency states, a signature of Markovianity, while at the same time reduce the slope of the distribution function at very low frequency (see inset). The slope allows us  to  extract an effective temperature $T_{\rm eff}$ to characterize the effect of dephasing. If the system was in true thermal equilibrium, the quantum fluctuation-dissipation theorem (FDT) would constrain the Keldysh and retarded components to obey the relation
\begin{align} \label{eq:FDT}
     1-2F_{\rm eq}(\omega) = \rm{tanh}\left( \frac{\beta \omega}{2}\right)
\end{align}
where \( \beta = 1/T \) is the inverse temperature of the system and $\mu=0$ in this particle-hole symmetric case. In the low-frequency limit, \( \omega \ll T \), the equilibrium distribution function behaves as \( 1-2F_{\text{eq}}(\omega) \sim \omega/2T \) allowing us to extract the temperature from a linear fit. 
\\
In a non-equilibrium system on the contrary there is no well-defined temperature and the FDT does not hold in general. Nonetheless, it is useful to use the left-hand
side of the FDT relation in Eq.~\ref{eq:FDT} to define an effective distribution function and extracting an effective temperature $T_{\rm eff}$.
We plot $T_{\rm eff}$  in Fig.~\ref{fig:spectralfunction}(c), where we compare the effect of charge and spin dephasing. We see that  $T_{\rm eff}$ increases with $\gamma$ as expected, yet the rate is significantly smaller for the charge dephasing case, leading to $T_{\rm eff}<T_K$ for all values of $\gamma_{\rm charge}$ considered in this work. On the other hand, the heating is parametrically stronger when dephasing acts on the spin degree of freedom, as seen from the effective temperature whose behavior at small $\gamma_{\rm spin}$ seems sub-linear growing above $T_K$ already for $\gamma_{\rm spin}\sim \Gamma $ (see Ref.~\cite{supplementary} for further data on the distribution function). This can be understood since the effective temperature probes heating of low energy degrees of freedom which are mainly spin fluctuations of the dot due to Kondo effect. The effective temperature depends also on the interaction $U$, particularly for the spin-dephasing case where the non-analytic behavior at small dephasing is enhanced, see Ref.~\cite{supplementary}. These results imply that monitoring of the charge degrees of freedom does not  immediately destroy the Kondo regime. We now discuss the consequences of this finding for charge transport by analyzing the conductance through the dot.

\begin{figure*}[t!] 
 \includegraphics[width=1.0\textwidth]{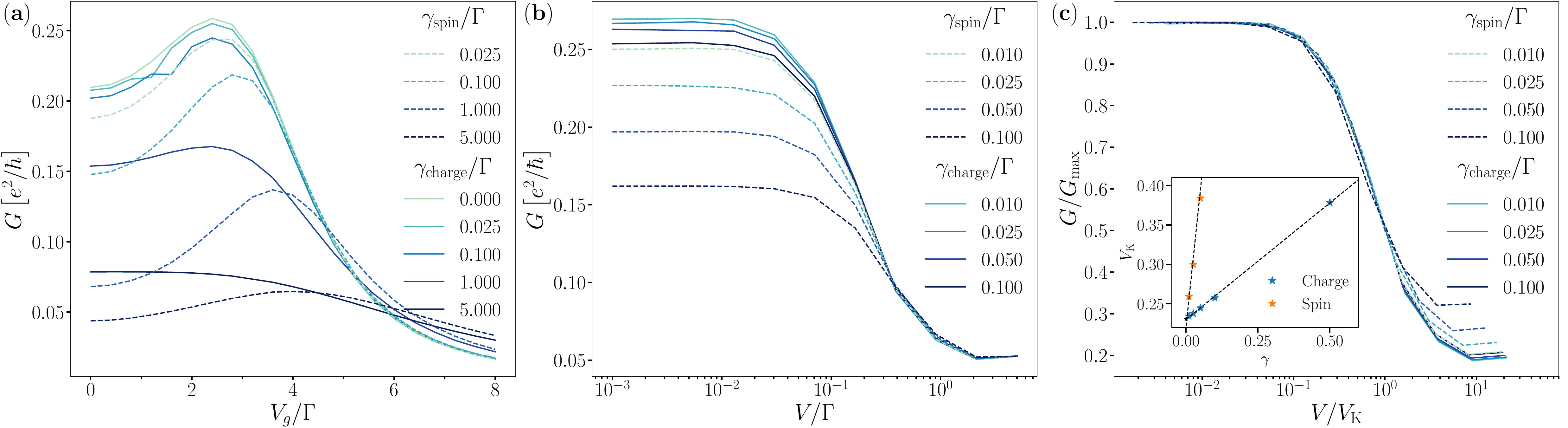}
 \caption{\label{fig:conductance}
Dissipative Anderson Impurity Model - (a) Differential conductance $G$ as a function of the gate voltage $V_g$ for increasing values of $\gamma_{\rm charge}$ and $\gamma_{\rm spin}$. (b) Differential conductance $G$ a function of the bias voltage for increasing values of $\gamma_{\rm charge}$ and $\gamma_{\rm spin}$. (c) The normalized conductance $G/G_{\rm max}$ displays a universal behavior as a function of the normalized bias $V/V_{K}$, highlighting the universality and scaling behavior across different charge and spin dephasing rate. Here, $G_{\rm max}$ denotes the zero-bias conductance at a given $\gamma$, and $V_{K}$ is defined as the bias voltage at which the conductance drops to half of the value it has with respect to $V=0$ for a given $\gamma$.}
\end{figure*}

\emph{Transport -- } We now focus on the particle transport through the quantum dot both in the linear and non-linear regime and discuss the effect of monitoring on current and conductance. The current operator is defined as the overall flux of particles flowing through the impurity at time $t$, namely $\m{I}(t) = \frac{1}{2}\left( \m{I}_{L}(t)  - \m{I}_R(t)  \right)$ with 
$\m{I}_\alpha=-e \frac{d N_\alpha}{dt}$  the instantaneous current flowing out from
lead $\alpha$ and $N_\alpha = \sum_{\bb{k}\sigma} c_{\bb{k}\sigma,\alpha}^\dagger c_{\bb{k}\sigma,\alpha} $ the number of conducting electrons in lead $\alpha$. The commutator can be 
evaluated explicitly due to the fact that the dissipation act only at the level of the impurity, leading to $\am{I}_\alpha (t)=\frac{i e}{\hbar} \sum_{\bb{k}\sigma} V_{\bb{k}\alpha} \langle d_\sigma^\dagger (t) c_{\bb{k}\sigma,\alpha}(t) - c_{\bb{k}\sigma,\alpha}^\dagger (t) d_\sigma (t) \rangle$.
For this reason the expression for the current is the same as for a non-dissipative dot and acquires the Meir-Wingreen form,~\cite{landauer1992meir,ferreira2023exact} (see Ref.~\cite{supplementary} for a proof) which in the case of equal lead self energies $\Gamma_R(\omega)=\Gamma_L(\omega)\equiv 2 \Gamma(\omega)$ becomes
\begin{align}\label{eqn:currentMWformula}
\langle \mathcal{I}\rangle=\frac{2e}{\hbar}\int \frac{d\omega}{2\pi}\Gamma(\omega)\left(F_{\rm R}(\omega)-F_{\rm L}(\omega)\right)\mbox{Im} \left[ - G^R(\omega) \right] \;.
\end{align}
Here, $F_{\alpha}(\omega) = F_{\rm eq}(\omega-\mu_\alpha)$ denotes the Fermi distribution of lead $\alpha$ with  chemical potential $\mu_\alpha$.  

We start discussing the linear conductance, $G=\partial I/\partial V_{V=0}$ and its dependence on the gate voltage $V_g$ and the dephasing rate $\gamma$. 
In Fig.~\ref{fig:conductance} (a) we plot $G(V_g)$ for different values of the charge $\gamma_{\rm charge}$ and spin $\gamma_{\rm spin}$ dephasing rates. In the standard (unitary) AIM conductance shows clear signatures of Kondo effect as well as Coulomb blockade physics. Here we see that for $\gamma_{\rm charge}=0$ the conductance has a peak at $V_g\sim U/4$ and a dip at the symmetric point due to finite temperature effects. As we switch on the dephasing in either spin or charge sector we see different behaviors: charge dephasing does not affect transport until a value $\gamma_{\rm charge}\sim \Gamma$ is reached, above which the suppression of coherent transport is clear. The spin dephasing on the other hand is very effective in reducing the conductance, particularly at $V_g\simeq0$, leaving only a broad peak at $V_g\sim U/2$. 

We then move to discuss the non-linear transport at finite voltage. In Fig.~\ref{fig:conductance} (b) we plot the conductance as a function of voltage for different types of dephasing. We see that the qualitative behavior is similar, namely dephasing reduces uniformly the low-bias conductance for $V\ll \Gamma$. Again, compared to the charge dephasing case the spin dephasing has a stronger effect on the suppression of transport. For voltages of the order $V\sim \Gamma $ all the $\gamma$ dependence disappears. Quite remarkably, as we see in Fig.~\ref{fig:conductance} (c),  we can rescale all the dephasing dependence and obtain an excellent scaling collapse of the normalized conductance $G/G_{\rm max}$. This is a remarkable result for a dissipative quantum impurity model, which suggests the existence of a universal scale $V_K$ controlling the dephasing problem, analog to the Kondo temperature. Indeed, for the unitary AIM the conductance is known to display universal scaling with respect to $T/T_K$ and $V/T_K$~\cite{we.lo.23}. The dependence of the crossover scale $V_K$, shown in the inset of Fig.~\ref{fig:conductance} (c), implies that the scaling collapse survives over a broader range of parameters for charge dephasing, in agreement with our overall picture that this process does not directly affect Kondo physics.

\emph{Conclusions --} In this work we have studied transport through an interacting 
monitored quantum dot modeled as a dissipative Anderson impurity model. We have discussed in particular how dissipation due to monitoring interplays with the physics of the Kondo effect, considering the separate effects of charge and spin monitoring on both spectral and transport properties of the model. Our results highlight the robustness of Kondo physics to charge monitoring, while spin dephasing turns out to destroy rapidly any signatures of the correlated state. We have shown that this difference can be understood in terms of local heating which remains moderate when the charge of the dot is monitored, while rapidly brings the system away from the Kondo regime when acting on the spin. The robustness of Kondo physics to charge dephasing is strikingly demonstrated in the scaling collapse of the non-linear conductance with respect to dephasing, suggesting that a form Kondo universality survives even in presence of dissipation. Our work suggests a number of interesting extensions, notably to heat transport where one can expect measurements to play a role in cooling~\cite{barontini2025quantum} and to non-reciprocal effects due to asymmetric couplings~\cite{ferreira2023exact}.

\emph{Acknowledgements --} We acknowledge computational resources from the Coll\'ege de France IPH cluster, the TU Graz A-Cluster, and the  Austrian Scientific Computing (ASC) infrastructure.
This research was funded by the Austrian Science Fund (FWF) [Grant DOI:10.55776/P33165],  by NaWi Graz and by the European Research Council (ERC) under the European Union's Horizon 2020 research and innovation programme (Grant agreement No. 101002955 -- CONQUER).  For the purpose of open access, the author has applied a CC BY public copyright licence to any Author Accepted Manuscript version arising from this submission. 

\bibliography{dissipativeAIM}

\pagebreak
\clearpage
\onecolumngrid
\begin{center}
		\large{\bf Supplemental Material to `Anderson Impurity Model with Dephasing: Steady-State and Transport'\\}
	\end{center}
	
\renewcommand{\thefigure}{S\arabic{figure}}

\newcounter{ssection}
\stepcounter{ssection}

\setcounter{table}{0}
\setcounter{page}{1}
\setcounter{figure}{0}
\setcounter{equation}{0}

\makeatletter
\renewcommand{\theequation}{S\arabic{equation}}

\appendix

In this Supplemental Material, we discuss:
\begin{enumerate}
    \item A Derivation of the Meir-Wingreen Formula
    \item The Auxiliary Master Equation Approach
    \item Impurity Spectral Function and Distribution Function for the Dissipative Resonant Level Model, Spectral properties of the Anderson impurity model subjected to spin dephasing , Effect of the interaction U on the effective temperature
\end{enumerate}

\section{Derivation of the Meir–Wingreen formula}
\newcommand{\beq}{\begin{equation}}
\newcommand{\eeq}{\end{equation}}
Due to the fact that the dephasing only acts on the impurity, the Meir Wingreen formula for the current~\cite{landauer1992meir} holds in the present case as well.
For completeness we provide here a derivation closely following Ref.~\cite{landauer1992meir}.
The current operator (from left to right) is defined as the overall flux of particles flowing through the impurity at time t, namely
\begin{align}
\label{it}
    \m{I}(t) = \frac{1}{2} \left[ \m{I}_{\rm L} - \m{I}_{\rm R}\right]
\end{align}
where $\rm{L/R}$ refer to the left/right metallic lead and $\m{I}_{\alpha}$ is the instantaneous current flowing out of lead $\alpha$,
\begin{align} \label{eqn:CurrentDef}
    \m{I}_\alpha (t) = - e \frac{d N_\alpha}{dt}
\end{align}
where $N_\alpha= \sum_{\bb{k}\sigma} c_{\bb{k},\sigma\alpha}^\dagger c_{\bb{k},\sigma \alpha}$ is the number of conducting electrons in lead $\alpha$ at time t. If one takes the trace over the density matrix evolving with the Lindblad master equation given in the main text it follows that the average current from lead $\alpha$ reads

\begin{align}\label{eq:currentalpha}
    \langle \m{I}_\alpha \rangle_t = \frac{ie}{\hbar} \sum_{\bb{k}\sigma}  \left(  V_{\bb{k} \alpha}^\ast  \langle d_\sigma^\dagger   c_{\bb{k},\sigma \alpha} \rangle_t  -  V_{\bb{k} \alpha} \langle c_{\bb{k},\sigma \alpha}^\dagger  d_\sigma \rangle_t \right)
\end{align}

All operators in Eq.~\eqref{eq:currentalpha} evolve in real time according to the Lindblad master equation.
This result follows since the jump operators only act on the impurity.

In the stationary state, the expectation value of the current operator simplifies to
(we now take $ V_{\bb{k} \alpha}$  to be real for simplicity)
\begin{align}
\label{ia}
    \langle \m{I}_\alpha \rangle &=  -i \frac{e}{\hbar} \sum_{\bb{k}\sigma}  \left(  V_{\bb{k} \alpha}  \langle c_{\bb{k},\sigma \alpha}^\dagger  d_\sigma \rangle - c.c.\right) \notag \\ & =  -\frac{e}{2 \hbar} \sum_{\bb{k}\sigma} \int \frac{ d \omega}{2\pi} \left[ V_{\bb{k} \alpha} G^{K}_{imp,\bb{k}\alpha}(\omega) +c.c.  \right] 
\end{align}
where in the second line we have used the definition of the  Green's function 
(which we take spin-independent), $G^{K}_{\rm imp, \bb{k}\alpha}(t,t^\prime) = -i \langle [d_\sigma(t),  c_{\bb{k}\sigma\alpha}^\dagger(t^\prime)] \rangle  $ connecting the lead and impurity degrees of freedom.
Since the leads are noninteracting 
one can use the Dyson equation with Langreth rules to obtain 
\beq
G^{K}_{imp,\bb{k}\alpha}(\omega) = V_{\bb{k} \alpha} \left[ G^R(\omega) g^K_{\bb{k}\alpha}  +   G^K(\omega)    g^A_{\bb{k}\alpha} \right]
\eeq
Since the local $G^K(\omega)$ are purely imaginary, 
we can rewrite  \eqref{ia} as
\begin{align}
\label{ia2}
 \langle \m{I}_\alpha \rangle &= 
- \frac{e}{ \hbar}  \int \frac{ d \omega}{2\pi} 
\sum_{\bb{k}\sigma}  V_{\bb{k} \alpha}^2 \left(-  \Im \left[ G^R(\omega) \right] \Im \left[ g^K_{\bb{k}\alpha} \right] +   \Im \left[ G^K(\omega) \right] \Im\left[ g^R_{\bb{k}\alpha} \right]  \right)   
\notag \\ & =
-\frac{e}{ \hbar}  \int \frac{ d \omega}{2\pi} 
\left( -\Gamma_\alpha(\omega) \Im \left[ G^K(\omega) \right]  + 2 \Im \left[ G^R(\omega) \right] \Gamma_\alpha(\omega) \left(1-2 F_{\alpha}(\omega) \right) \right)
\end{align}
where we have used

\beq
\sum_{\bb{k}\sigma}  V_{\bb{k} \alpha}^2 \Im g^R_{\bb{k}\alpha} = -2 \Gamma_\alpha(\omega)  \quad\quad 
g^K_{\bb{k}\alpha} = 2 i \Im \left[ g^R_{\bb{k}\alpha} \right] \left(1-2 F_{\alpha}(\omega) \right)
\eeq
Inserting \eqref{ia2} in \eqref{it} yields
\beq
\label{mit2}
  \am{I}(t) = -\frac{2 e}{ \hbar}  \int \frac{ d \omega}{2\pi}  
\left(
   -\frac12 \left(\Gamma_L(\omega) -\Gamma_R(\omega) \right) \left(\Im  \left[ G^K(\omega) \right] -2 \Im \left[ G^R(\omega)  \right] \right)  
- 2 \Im \left[ G^R(\omega) \right] \left( \Gamma_L(\omega) F_{L}(\omega) - \Gamma_R(\omega) F_{R}(\omega) \right)
\right)
\eeq
We can use
\beq
\Im \left[ G^K(\omega) \right]  = 2 \Im  \left[ G^<(\omega) \right] +  2 \Im \left[ G^R(\omega) \right]
\eeq
to rewrite \eqref{mit2} as
\beq
  \am{I}(t) = -\frac{2 e}{ \hbar}  \int \frac{ d \omega}{2\pi}  
\left\{
  \left( \Gamma_R(\omega) -\Gamma_L(\omega) \right) \Im \left[ G^<(\omega)  \right]   
+ 2 \left( \Gamma_R(\omega) F_{R}(\omega) - \Gamma_L(\omega) F_{L}(\omega) \right) \Im \left[ G^R(\omega)   \right] 
\right\}
\eeq

As pointed out in Ref.~\cite{landauer1992meir}, 
in an equilibrium situation $F_{R}(\omega)=F_{L}(\omega)$ and 
$\Im \left[ G^<(\omega) \right]  = -2 \Im \left[ G^R(\omega) \right] F_{R/L}(\omega)$ so that the current vanishes. Due to the dephasing, however, the second condition is not fullfilled even in the case in which the first one is, i.e. leads in equilibrium. As pointed out in Ref.~\cite{ferreira2023exact}, this can produce a nonreciprocal current flowing without bias under the absence of certain symmetry conditions. This is because the dephasing term can be seen also as a coupling to a bosonic bath at infinite temperature.

\section{Auxiliary Master Equation Approach}

The auxiliary master equation approach (AMEA)~\cite{ar.kn.13,do.nu.14} is a non-equilibrium steady-state impurity solver and in the following we will give a brief overview. For details the reader is referred to Refs.~\cite{do.nu.14,ar.do.18,we.lo.23}

A correlated impurity problem is completely described by the hybridization function, which defines the uncorrelated fermionic reservoirs, the on-site energy and the Hubbard interaction. Within AMEA, the ``physical'' fermionic reservoirs are mapped onto an ``auxiliary'' open system, which consists of a finite number of bath sites, $N_B$, that are in turn coupled to Markovian environments described by linear Lindblad terms. 
The Lindblad and hopping parameters of the auxiliary system are obtained by a fit to the physical hybridization function.
The accuracy of the mapping is determined solely by the difference between the (retarded and Keldysh components) of the hybridization functions of the physical and the auxiliary reservoirs. 
This difference decreases exponentially with increasing $N_B$,
as was shown Ref.~\cite{do.so.17}, and is already quite small for $N_B \gtrsim 6 $. 
This is due to the fact that the number of fit parameters is already quite large for this value of $N_B$.
For finite, not too large $N_B$ the auxiliary many body problem can be solved by exact methods such as Lanczos or Matrix product states, whereby the many-body density matrix has to be addresses. Concretely, one finds the steady state density matrix and the Green's functions directly in real frequencies.
 In this paper, we use a configuration interaction based approach~\cite{we.lo.23}, that uses Lanczos/Arnoldi for the computation of the steady state and the Green's functions, which has been further improved by introducing a functional interpolation scheme~\cite{we.ar.25}. Throughout this paper we use $N_\text{b} = 8$.
\section{Additional Results}
In this section we present additional results on the dissipative Anderson Impurity Model: (i) benchmark of the auxiliary master equation on the dissipative resonant level model. (ii) Spectral function and distribution function for the spin dephasing case. (iii) The effect of the interaction U on the effective temperature.

\subsection{Impurity Spectral Function and Distribution Function for the Dissipative Resonant Level Model}
\begin{figure}[!t] 
 \includegraphics[width=1.0\textwidth]{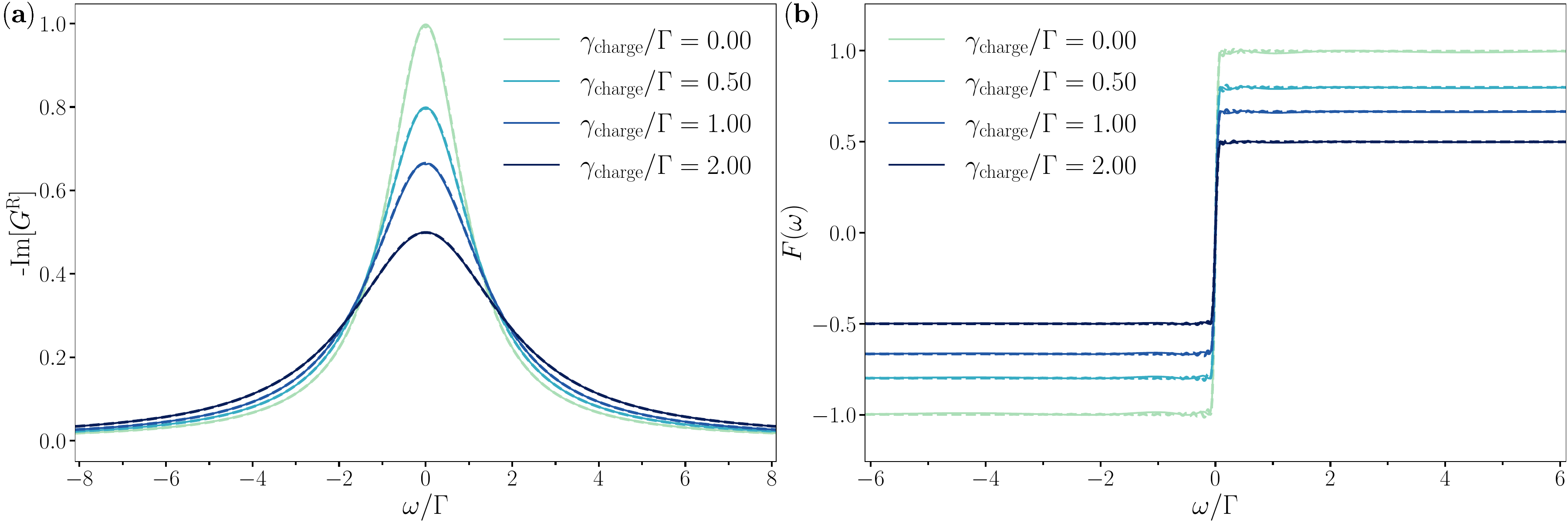}
    \caption{\label{fig:noninteractingexact} Dissipative Resonant Level model with charge dephasing $L = \sqrt{\gamma_{\rm charge}} \sum_\sigma n_\sigma$. (a) Impurity spectral function for the half-filled dissipative resonant level
model, corresponding to $U = 0$, and increasing values of $\gamma_{\rm charge}$ at fixed temperature $T=0.02 \Gamma $. (b) Corresponding distribution function $F(\omega)$ . The solid lines represent results from the auxiliary master equation approach, while the dotted lines show the exact solution obtained via the Keldysh formalism.   }
\end{figure}

We now consider the dissipative but non-interacting case, with U=0, corresponding to a dissipative Resonant Level Model. Due to the quadratic nature of the quantum jump, we stress that the system is still interacting, i.e non-gaussian in the dissipative way. However the specific nature of the charge and spin dephasing make it possible to close exactly the equations of motions for the retarded Green's function in terms of a Dyson equation.~\cite{ji.fe.22} We obtain therefore in the real time domain,
\begin{align}
    \Sigma (t,t^\prime) = \gamma_{\alpha} \delta(t,t^\prime) G(t,t^\prime) 
\end{align}
In Fig.~(\ref{fig:noninteractingexact}), we plot the imaginary part of the impurity retarded green's function (left panel) and the corresponding distribution function obtained with the Auxiliary mas ter equation in the $U=0$ case for increasing value of the charge dephasing rate. A comparaison with the exact solution obtained from the Keldysh formalism is shown in dashed lines, we see an essentially perfect agreement between the Auxiliary master equation approach and the exact result. We stress that this comparison is non trivial and, for example, it fails for smaller number of bath sites in the AMEA approach.

As $\gamma_{\rm charge}$ increases,  the resonance at the Fermi level is broadened and at the same time the value at $\omega=0$ decreases with $\gamma_{\rm charge}$, which is a signature of loss of coherence due to the Markovian environment.  At the same time, the distribution function deviates from the equilibrium Fermi-Dirac form, reflecting the fact that the system is driven out of equilibrium and relaxes to a non-thermal state. Importantly, the dissipation acts uniformly across all frequencies, meaning that the dissipation do not discriminate between low and high-energy excitations, it affects all many-body states with equal strength, regardless of their energy. 

In the non-interacting Anderson impurity model at half-filling, all states are dynamically equiprobable, and the dissipation does not preferentially target low or high-energy excitations. As a result, the distribution function becomes progressively flatter with increasing dephasing, reflecting a uniform smearing of occupations across the spectrum. This energy-insensitive behavior is typical of dephasing processes, which tends to push the system toward a fully mixed state, similar to an infinite-temperature.

\subsection{Spectral Properties of the Anderson Impurity Model Subjected to Spin Dephasing}
\begin{figure}[!t] 
 \includegraphics[width=1.0\textwidth]{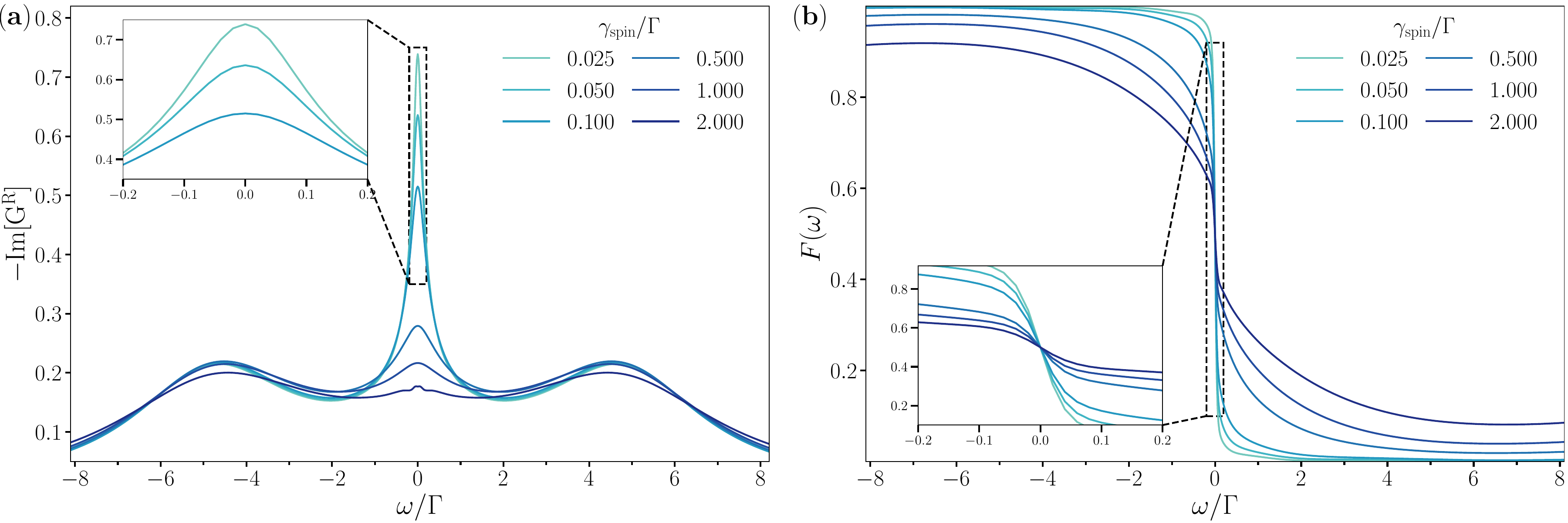}
    \caption{\label{fig:SpinDephasing_spectralDistribu} Anderson Impurity Model with spin dephasing $L= \sqrt{\gamma_{\rm spin}} \sum_\sigma  \sigma n_\sigma $. (a) Impurity spectral function at half-filling for interaction strength $U = -2\epsilon_d = 8\Gamma$, shown for increasing values of the charge dephasing rate $\gamma_{\rm spin}$.
    (b) Corresponding distribution function $F(\omega)$ as $\gamma_{\rm spin}$ increases. }
\end{figure}
In addition to the charge dephasing case discussed in the main text, we present here further results concerning the spectral properties of the Anderson impurity model subjected to spin dephasing, which we recall is implemented via the quantum jump operator
\begin{align}
    L_{\rm spin} = \sqrt{\gamma_{\rm spin}} \sum_\sigma \sigma n_\sigma
\end{align}
In Fig.~(\ref{fig:SpinDephasing_spectralDistribu}), we show the imaginary part of the Green's function and the corresponding distribution function for increasing values of the spin dephasing rate $\gamma_{\rm spin}$. As $\gamma_{\rm spin}$ increases, we observe a pronounced suppression of the Kondo resonance in the spectral function. This behavior signals the breakdown of coherent spin fluctuations that are essential for the formation of the Kondo singlet. Notably, while the sharp Kondo peak is rapidly suppressed, the high-energy Hubbard bands remain largely unaffected by spin dephasing, indicating that charge excitations are more robust against this type of dephasing. The contrasting behavior between the Kondo peak and the Hubbard bands highlights that spin dephasing acts primarily on the low-energy spin degrees of freedom without significantly perturbing the charge sector~\cite{vanhoecke2025kondozenocrossoverdynamicsmonitored}. 

This interpretation is further supported by the behavior of the distribution function. In contrast to charge dephasing, whose effect tends to be more frequency-independent, spin dephasing predominantly alters the distribution function in the low-frequency region, corresponding to the energy window where spin fluctuations dominate. This selective suppression around $\omega \approx 0$ is consistent with the loss of Kondo coherence. On the other hand, the distribution remains almost unchanged at high frequencies $|\omega| \gg 0$, where charge fluctuations prevail, further confirming that spin dephasing acts mainly on the spin degrees of freedom and effectively suppresses Kondo correlations without significantly affecting the charge fluctuation regime.

\subsection{Effect of the Interaction U on the Effective Temperature}
\begin{figure}[!t] 
 \includegraphics[width=1.0\textwidth]{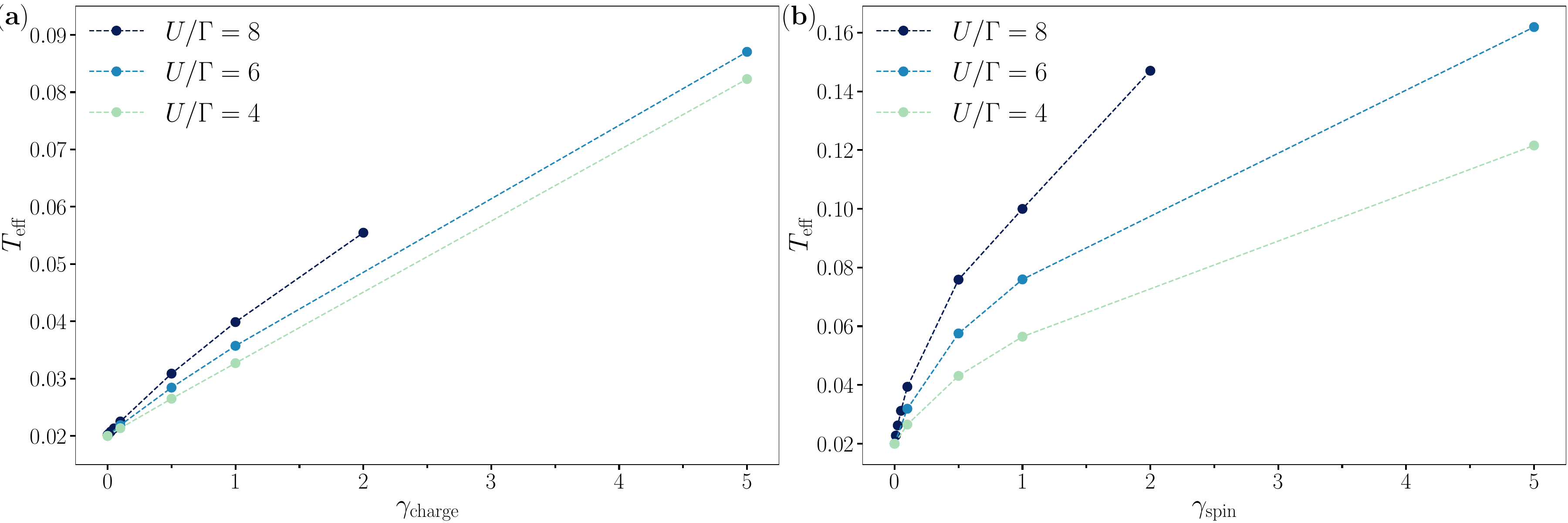}\caption{\label{fig:effectivetemperature_vs_U} Dissipative Anderson Impurity Model - Effective temperature $T_{\rm eff}$ of the dissipative Anderson impurity model as a function of the dissipation strength, for various values of the interaction U.  The left panel corresponds to charge dephasing, while the right panel shows the case of spin dephasing. The leads are held at a finite temperature $T=0.02 \Gamma$, and are modeled with a flat (wide-band) density of states. }
\end{figure}

In this section, we provide additional results on the effective temperature $T_{\rm eff}$ extracted from the distribution function $F(\omega)$. Specifically, we discuss the influence of the Coulomb interaction $U$ on $T_{\rm eff}$ for both the charge and spin dephasing. In Fig.~(\ref{fig:effectivetemperature_vs_U}), we plot the effective temperature $T_{\rm eff}$ as a function of dephasing rate $\gamma_{\alpha}$ for different values of interaction $U$. The left panel corresponds to charge dephasing ($b = charge$) and the right panel to spin dephasing ($b = spin$). We see that in both cases interaction leads to an increase of the effective temperature, an effect which is particularly evident for spin-dephasing where increasing $U$ makes $T_{\rm eff}$ almost non-analytic at small $\gamma$, as if heating in the spin dephasing case was controlled by a non-perturbative effect. We can understand the strong dependence of heating from interaction in the spin dephasing case by noticing that $T_{\rm eff}$ refers to the low-energy degrees of freedom, which are particularly sensitive to spin fluctuations in the strong interacting regime. 

\end{document}